\documentclass[lettersize,journal]{IEEEtran}
\usepackage{amsmath,amsfonts}
\usepackage{algorithmic}
\usepackage{algorithm}
\usepackage{array}
\usepackage[caption=false,font=normalsize,labelfont=sf,textfont=sf]{subfig}
\usepackage{textcomp}
\usepackage{stfloats}
\usepackage{url}
\usepackage{verbatim}
\usepackage{graphicx}
\usepackage{cite}
\usepackage{amsmath,amssymb,amsfonts}
\usepackage{multirow}
\usepackage{makecell}
\usepackage{float}
\usepackage{amssymb}
\usepackage{amsmath}
\usepackage{hyperref}
\usepackage[normalem]{ulem}
\useunder{\uline}{\ul}{}
\begin{document}

\renewcommand{\arraystretch}{1.1}

\title{Optimized Vessel Segmentation: A Structure-Agnostic Approach with Small Vessel Enhancement and Morphological Correction}

\author{Dongning Song, Weijian Huang, Jiarun Liu, Md Jahidul Islam, Hao Yang and Shanshan Wang*
\thanks{This research was partly supported by the National Natural Science Foundation of China (No. 62222118, No. U22A2040), Shenzhen Science and Technology Program (No. RCYX20210706092104034, No. JCYJ20220531100213029), Shenzhen Medical Research Fund (No. B2402047), the Major Key Project of PCL under Grant PCL2024A06, Key Laboratory for Magnetic Resonance and Multimodality Imaging of Guangdong Province (No. 2023B1212060052), and Youth lnnovation Promotion Association CAS.

Thanks, D. Song, W. Huang, J. Liu, M. J. Islam, H. Yang, and S. Wang are with the Paul C. Lauterbur Research Center for Biomedical Imaging, Shenzhen Institutes of Advanced Technology, Chinese Academy of Sciences, Shenzhen 518055, China (e-mail: ss.wang@siat.ac.cn).}}


% \IEEEpubidadjcol  % 调整出版物识符位置
% Remember, if you use this, you must call \IEEEpubidadjcol in the second

\maketitle

\begin{abstract}
Accurate segmentation of blood vessels is essential for various clinical assessments and postoperative analyses. However, the inherent challenges of vascular imaging—such as sparsity, fine granularity, low contrast, data distribution variability, and the critical need for preserving topological structure—make generalized vessel segmentation particularly complex. While specialized segmentation methods have been developed for specific anatomical regions, their over-reliance on tailored models hinders broader applicability and generalization. General-purpose segmentation models introduced in medical imaging often fail to address critical vascular characteristics, including the connectivity of segmentation results.
To overcome these limitations, we propose an optimized vessel segmentation framework: a structure-agnostic approach incorporating small vessel enhancement and morphological correction for multi-modality vessel segmentation. To train and validate this framework, we compiled a comprehensive multi-modality dataset spanning 17 datasets and benchmarked our model against six SAM-based methods and 17 expert models. The results demonstrate that our approach achieves superior segmentation accuracy, generalization, and a 34.6\% improvement in connectivity, underscoring its clinical potential. An ablation study further validates the effectiveness of the proposed improvements. We will release the code and dataset at \href{https://github.com/Hk416mod2/OVS-Net} {github.com} following the publication of this work.
\end{abstract}


\section{Introduction}
\IEEEPARstart{V}{essel} segmentation plays a crucial role in analyzing various related diseases\cite{mean_vessel_seg}. 
%For example, changes in structural parameters such as the diameter, branching angle, and curvature of retinal vessels can aid in diagnosing fundus diseases like diabetic retinopathy\cite{retina1} and arteriosclerosis\cite{retina2}. Quantitative analysis of coronary angiography is commonly used to assess myocardial infarction and coronary atherosclerosis\cite{ca1}, while automatic segmentation of coronary arteries in X-ray angiography assists in analyzing coronary artery disease\cite{DSA}. 
However, manual observation by doctors is often time-consuming, tedious, and requires a high level of expertise\cite{doctor_hard}. Therefore, developing an automatic, accurate, and highly generalizable vessel segmentation method using computer technology holds significant promise for clinical applications.

In medical image analysis, many pioneers have developed automated vessel segmentation models to reduce clinical workloads and enhance diagnostic efficiency. Early research primarily focused on traditional machine learning algorithms, in which vessels were identified and segmented using manually extracted features. For instance, Chaudhuri \textit{et al.}\cite{retinaseg1989} proposed a method utilizing a two-dimensional matched filter for retinal vessel detection, while Hoover \textit{et al.}\cite{hoover2000locating} employed matched filters for vessel segmentation.
Recently, with the advancement of deep learning—particularly the development of attention mechanisms\cite{ViT} and convolutional neural networks (CNNs)\cite{alexnet}—automated vessel segmentation has seen substantial improvements in both efficiency and accuracy. These innovations have further strengthened the role of computer-assisted medical analysis. Building on these advancements, recent progress in artificial intelligence has led to the emergence of large-scale foundational models\cite{fundationmodel1}, ushering in a new era of AI with remarkable zero-shot and few-shot generalization capabilities across various downstream tasks. Among these foundational models, the Segment Anything Model (SAM)\cite{SAM} has gained attention as a universal model for visual segmentation. Its robust cross-object segmentation abilities and impressive zero-shot generalization have made SAM particularly promising in medical imaging applications, demonstrating significant potential in addressing challenges traditionally faced by task-specific segmentation models\cite{samformed?}.

\begin{figure}[!t]
\centering
\includegraphics[width=0.5\textwidth]{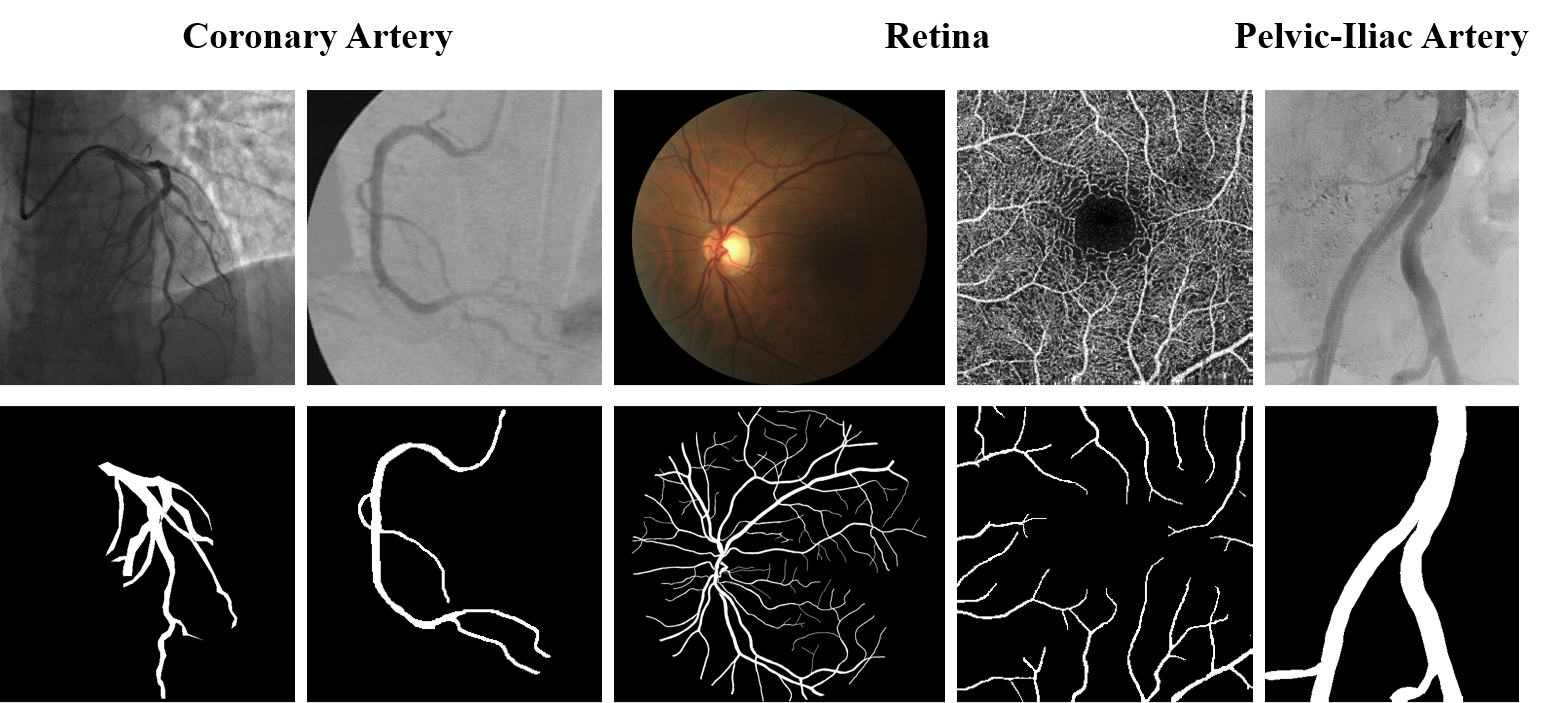}

\caption{The distribution of vascular imaging data shows significant variation across different anatomical regions and acquisition protocols, posing challenges for the generalization of task-specific expert models.}
\label{fig1}
\end{figure}

% Although various deep learning models based on convolution\cite{U-Net} and attention\cite{ViT} mechanisms have achieved accurate segmentation of vessel images\cite{vessel_seg_work1}, a significant limitation of many current medical image segmentation models is their task specificity. When deploying a trained vessel segmentation model directly on imaging data from different anatomical regions or protocols, the network's performance may experience a significant decline. This lack of generalizability greatly hinders the widespread application of vessel segmentation models in clinical practice.
% With the rapid advancement of artificial intelligence technology in recent years, the emergence of large-scale foundational models\cite{fundationmodel1} has fundamentally transformed the field of artificial intelligence and ushered in a new era, owing to their remarkable generalization capabilities across a wide range of downstream tasks. 
% As a general-purpose foundational model for visual segmentation, the Segment Anything Model (SAM)\cite{SAM} has demonstrated unprecedented potential in medical imaging due to its exceptional ability to segment across diverse objects and its robust generalization capabilities\cite{MedSAM, MSA}. However, despite the impressive performance of the SAM model in the medical field, there is still no universal segmentation model specifically designed for vascular imaging, and existing segmentation methods exhibit limitations in maintaining topological structures.

However, accurate, generalized, and topology-preserving vessel segmentation remains a challenging task. While blood vessels are inherently difficult to segment due to their fine granularity, sparsity, low contrast, and variable shapes, the key challenges lie in the model's generalization across diverse datasets and its ability to preserve topological integrity during segmentation.
First, as shown in fig. \ref{fig1}, significant variations in vascular imaging data across different anatomical locations and acquisition protocols\cite{domain_gap} create substantial difficulties for model generalization. Models trained on a single dataset or modality often fail to perform well when applied to new datasets, leading to resource inefficiency, redundant model development, and deployment obstacles. The ability to generalize across such diverse data distributions is critical to ensuring the practical applicability of vessel segmentation models in real-world clinical settings.
Second, unlike other segmentation tasks, vessel segmentation requires a strong focus on preserving the topological structure of the vascular network. Inaccuracies such as discontinuities, fractures, or merging of parallel vessels can severely impact the clinical utility of the segmentation\cite{dvaerefiner}. For example, errors at bifurcations or crossings in retinal vessels, which are often tiny and curved, can lead to incorrect analyses, potentially affecting surgical planning, lesion diagnosis, and interventional treatments\cite{cldice}. Moreover, segmentation algorithms optimized for overlap scores tend to overlook small or fragile vascular structures, which can result in fragmented segmentation, especially problematic in clinical applications where accuracy is paramount.
Thus, to achieve accurate, generalized, and topology-preserving vessel segmentation, models must address both the challenges of generalization across varying data distributions and the critical need for topological integrity in the segmentation process.

In this paper, we propose OVS-Net, a structure-agnostic approach designed for optimized vessel segmentation.
To adapt the universal segmentation model for vessel segmentation, we developed a dual-branch feature extraction module, consisting of a macro vessel extraction module and a micro vessel enhancement module, and modified the decoder to improve vessel segmentation mask prediction. Specifically, the original encoder is tasked with extracting features from macro and main vessels, while the micro-vessel enhancement module amplifies features of smaller vessels. Additionally, lower-level features of the micro-vessels are integrated into the decoder to further exploit multi-scale vascular information.
Furthermore, we designed a topology repair network to address disconnected vessels resulting from the segmentation process, making the overall architecture more suitable for clinical use. To train and evaluate our model, we compiled a large vascular dataset incorporating 17 datasets, extensive experiments demonstrated the effectiveness of our method. 
The key contributions of this paper are summarized as follows:

% Specifically, OVS-Net comprises four main modules: the Global Information Extraction Module, the micro vessel enhancement module, the Pre-Segmentation Module, and the Morphological Correction-based post-processing network. Considering the sparsity of vessels and the wide receptive field of attention networks, the Global Information Extraction Module adopts the Vision Transformer (ViT) image encoder from SAM. This serves two purposes: leveraging the pre-trained knowledge of the SAM model and adapting it to vessel scenarios by designing feature and spatial adapters for the ViT image encoder. For the micro vessel enhancement module, we employ a convolutional network based on ConvNext and FPN to enhance the features of the micro-vessel. This module is connected to the global information extraction module via cross-attention, enabling the network to better learn information about different types of vessels. Additionally, we use SAM’s mask decoder as the Pre-Segmentation Network, augmenting it with feature connections. Finally, we incorporate a simple U-Net as the Morphological correction-based post-processing network to refine the segmentation mask generated by the Pre-segmentation network. This step connects broken vessel segmentations while preserving the original vascular connectivity, enhancing the network’s clinical applicability. The main contributions of this paper are summarized as follows:

\begin{enumerate}

% \item We developed OVS-Net, a structure-agnostic segmentation method for optimized vessel segmentation.

% \item We collected 17 datasets to train and evaluate our method, and provided the data splits and download instructions on \href{https://github.com/Hk416mod2/OVS-Net}{github.com}.

% \item To enhance the model's ability to effectively extract small vessel structures, we designed a macro vessel extraction-micro vessel enhancement dual-branch module. We also included feature connections into the mask decoder to better adapt universal segmentation method for vessel segmentation tasks better and additionally designed a post-processing network to boost connectivity and clinical applicability.

% \item Extensive testing on 17 datasets demonstrated that Our model obtained higher segmentation accuracy, generalization, and a 34.6\% improvement in connectivity, demonstrating the model’s substantial potential for clinical applications.

\item Development of OVS-Net: We introduced OVS-Net,  a structure-agnostic segmentation framework designed for optimized vessel segmentation across diverse modalities and anatomical structures.

\item Comprehensive Dataset Collection: We curated 17 datasets to train and evaluate OVS-Net, providing data splits and download instructions on GitHub to promote transparency and reproducibility.

\item Innovative Module Design: Developed a macro vessel extraction-micro vessel enhancement dual-branch module to effectively capture small vessel structures. Integrated feature connections into the mask decoder, adapting universal segmentation methods specifically for vessel segmentation tasks. Designed a post-processing network to improve segmentation connectivity and enhance clinical applicability.

\item Extensive Experimental Validation: Conducted rigorous testing across 17 datasets, demonstrating OVS-Net's superior segmentation accuracy, generalization capabilities, and a 34.6\% improvement in connectivity, showcasing its potential for clinical applications.

\end{enumerate}

\section{Related Work}

\subsection{Vessel Segmentation Methods}

Accurate segmentation of blood vessels is crucial for evaluating various disorders, particularly cardiovascular diseases, as well as for surgical planning and monitoring treatment progress. This process plays a vital role in clinical analysis. With advancements in deep learning, there has been growing recognition of the importance of integrating multi-scale information in vessel segmentation. As a result, numerous methods have been proposed for the automated segmentation of blood vessels.

In fundus retinal vessel segmentation, Wu \textit{et al.}\cite{SCS-Net} proposed SCS-Net, which combines a semantic context aggregation module to enhance multi-scale information extraction and an adaptive feature fusion module to improve integration across adjacent layers. Jiang \textit{et al.}\cite{Covi-Net} introduced Covi-Net, which integrates a Local and Global Feature Aggregation (LGFA) module, preserving the ability to process local information while facilitating the processing of distant global information. Kang \textit{et al.}\cite{CFI-Net} proposed CFI-Net, a model that employs joint refinement downsampling to mitigate spatial information loss caused by downsampling, preserving the details and edge features of retinal vessels.
In OCTA image segmentation, Ning \textit{et al.}\cite{FR-Net} proposed FR-Net, an efficient and accurate network that leverages a modified Recurrent ConvNeXt Block within a full-resolution convolutional framework. This network reduces parameters and accelerates inference speed. Ma \textit{et al.}\cite{CosNet} addressed data imbalance in the OCTA dataset by proposing CoSNet, which utilizes global contrastive learning to allow the network to learn both local and global features. Additionally, Ma \textit{et al.}\cite{OCTA-Net} proposed OCTA-Net, which uses a coarse segmentation module to generate preliminary vessel confidence maps and a fine segmentation module to refine the contours and shapes of retinal microvessels.
In coronary artery segmentation, Chang \textit{et al.}\cite{SE-RegUNet} proposed SE-RegUNet, which combines a RegNet encoder with squeeze-and-excitation blocks to enhance feature extraction. Liu \textit{et al.}\cite{FR-UNet} introduced FR-UNet, which uses a multi-resolution convolutional mechanism to expand both horizontally and vertically while maintaining full image resolution.
Despite their contributions, the aforementioned methods are specialized models tailored to specific datasets, and their performance may significantly degrade when applied to new data or different scenarios, which hinder deployment and limit their generalization to other vascular regions.

\subsection{Segment Anything Models}

In recent years, visual foundational models tailored for specific segmentation tasks have gained increasing attention. The Segment Anything Model (SAM)\cite{SAM}, the first large foundational model designed for segmentation, has not only demonstrated outstanding performance on natural images but also shown remarkable promise for generalization in medical image analysis. Huang \textit{et al.}\cite{samformed?} evaluated SAM's applicability in medical contexts, highlighting its significant potential in medical image segmentation. Ma \textit{et al.}\cite{MedSAM} proposed MedSAM, trained on a large dataset of over 1 million images across 11 imaging modalities, which outperformed U-Net expert models in segmentation tasks. SAM-med2d\cite{sm2d} and Medical SAM Adapter\cite{MSA} introduced Adapter modules, while SAMed\cite{SAMed} utilized LoRA for efficient fine-tuning, achieving high performance across multiple medical image segmentation tasks. Chai \textit{et al.}\cite{LadderNet} developed a staged fine-tuning approach that integrates a complementary CNN encoder with the SAM architecture, focusing on fine-tuning only the CNN and SAM decoder to minimize computational resources and training time. Similarly, Lin \textit{et al.}\cite{SAMUS} proposed SAMUS, which integrates a parallel CNN branch to inject local features into the ViT encoder, with position and feature adapters to modify SAM for clinically friendly deployment on smaller input sizes. Qin \textit{et al.}\cite{DB-SAM} introduced DB-SAM, combining a ViT branch and a convolution branch to bridge the domain gap between natural images and 2D/3D medical images.

Despite the impressive performance of SAM in general medical segmentation, its results in vascular imaging may not be as favorable. While SAM’s ViT-based image encoder excels at global context modeling and provides a larger receptive field, it tends to capture global and low-frequency information\cite{HQ-SAM}. However, vessel images are characterized by fine-grained and sparse features, with numerous micro-vessels, posing a challenge for the ViT encoder, which is less effective at capturing detailed, high-frequency information. While fine-tuning SAM on vascular data can improve performance, it often results in the loss of small vessels and fragmentation.

To address these issues, Ke \textit{et al.}\cite{HQ-SAM} proposed HQ-SAM, which introduced a learnable high-quality output token into the SAM mask decoder to predict more accurate masks. They also fused mask encoding with early and final ViT features to enhance mask details. Yuan \textit{et al.}\cite{RWKV-SAM} proposed RWKV-SAM, which combined convolution and RWKV operations with an efficient decoder and multi-scale tokens to generate high-quality segmentation masks. Despite these advancements, there is still no universal segmentation model specifically designed for vascular imaging. Furthermore, existing methods often overlook topological preservation in segmentation, limiting their effectiveness in universal vessel segmentation.

% Chai \textit{et al.}\cite{LadderNet} employed a staged fine-tuning approach, combining a complementary CNN encoder with the standard SAM architecture, focusing solely on fine-tuning the additional CNN and SAM decoder to reduce computational resources and training time. Similarly, Lin \textit{et al.} proposed SAMUS\cite{SAMUS}, which integrates a parallel CNN branch to inject local features into the ViT encoder, along with a position adapter and feature adapter to adjust SAM from large input sizes to smaller ones for more clinically friendly deployment. Ke \textit{et al.} introduced HQ-SAM\cite{HQ-SAM}, which designed a learnable high-quality output token to be injected into SAM's mask decoder, responsible for predicting high-quality masks and merging mask encoding with early and final ViT features to improve mask detail. Yuan \textit{et al.}proposed RWKV-SAM\cite{RWKV-SAM}, which incorporates a mixed backbone of convolutional and RWKV operations and a highly efficient decoder leveraging multi-scale tokens to achieve high-quality masks. Qin \textit{et al.} introduced DB-SAM\cite{DB-SAM}, which contains a ViT branch and a convolution branch, aiming to effectively bridge the domain gap between natural images and 2D/3D medical images. However, to our knowledge, there is currently no model specifically designed for universal vessel segmentation tasks and it remains a challenging problem.

\begin{figure*}[!t]
\centering
\includegraphics[width=\textwidth]{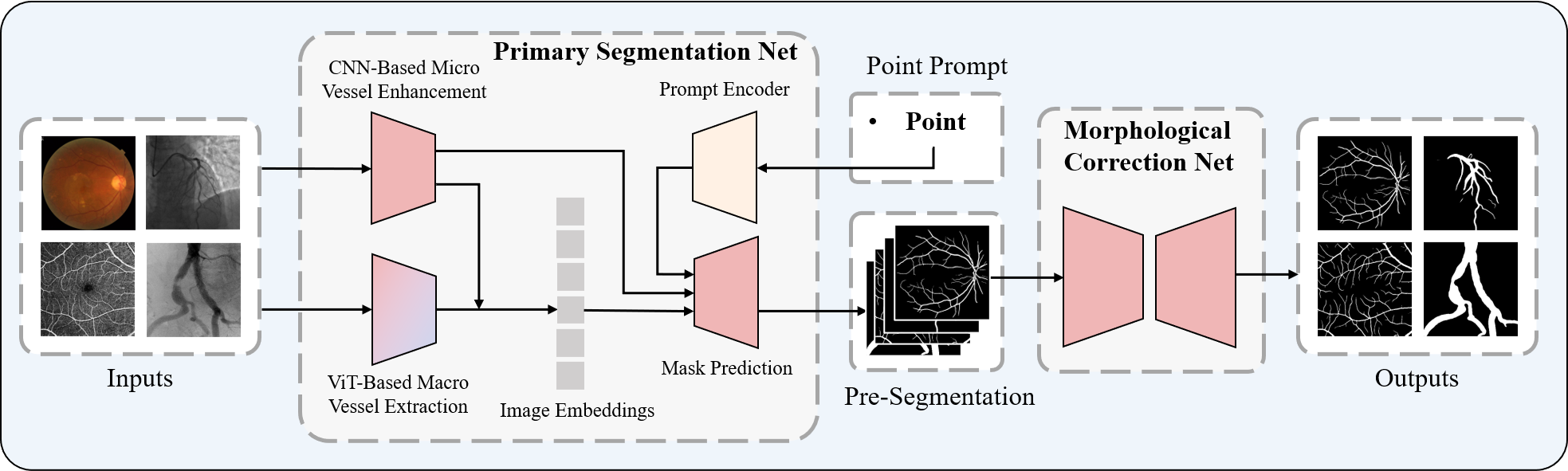}
\caption{Framework diagram of OVS-Net, which comprises five main modules: the ViT-based macro vessel extraction module, the CNN-based micro vessel enhancement module, the mask prediction module, the prompt encoder and the morphological correction-based post-processing network.}
\label{fig2}
\end{figure*}

\subsection{Topology Correction-Based Post-Processing Method}
The connectivity of vascular structure segmentation holds significant clinical value in medical applications\cite{CorSegRec, cldice}. Although it may not lead to a large improvement in overlap metrics such as the Dice score, it considerably boosts the practical interpretability and visual quality for physicians\cite{dvaerefiner}. The issue of connectivity differs from general segmentation tasks, since the area prone to disconnection normally arise in thin microstructure, which is less highlighted in overlap-oriented optimizations\cite{deepclosing}. As a consequence, common segmentation optimization algorithms mostly fail to handle this, resulting in vessel discontinuities. To address this challenge, various research has proposed additional post-processing strategies that enhance segmentation connectivity.

Teichmann \textit{et al.} proposed ConvCRF\cite{connfcrf}, which combines the structured modeling capabilities of Conditional Random Fields (CRF) with the feature extraction power of Convolutional Neural Networks (CNN) to effectively execute CRF on GPUs. Jha \textit{et al.} introduced a dual UNet architecture\cite{doubleunet}, where two UNets are employed to generate coarse and refined segmentation results, respectively. Ye\textit{et al.} designed VSR-Net\cite{vsr-net}, which utilizes a Curve Clustering Module (CCM) to build fragmented clusters, then patches the broken segments with a specially designed Curve Merging Module (CMM). Wang \textit{et al.} introduced SegRefiner\cite{segrefiner}, which incorporates a diffusion model to refine coarse masks by employing a discrete diffusion process.

\section{Method}

\subsection{Overview}

To adapt the universal segmentation model for vessel segmentation while maintaining its strong generalization capability and leveraging its pre-trained weights, we developed OVS-Net. The model consists of five main modules: the macro vessel extraction module, the micro vessel enhancement module, the mask prediction module, the prompt encoder module, and the morphological correction-based post-processing network module. The overall architecture is illustrated in Figure \ref{fig2}.
The macro vessel extraction module retains the ViT image encoder backbone from SAM and introduces feature and spatial adapters to better align the model with vascular characteristics. The micro vessel enhancement module uses a convolutional network based on ConvNext and FPN to improve the extraction of small vessel features. We kept the original design of the prompt encoder. The mask prediction module builds on the SAM mask decoder backbone, with enhanced feature connections to improve segmentation accuracy. Finally, the morphological correction-based post-processing network module is based on a simple U-Net architecture, designed to refine the pre-segmentation mask, ensuring improved connectivity and structural integrity in the final output.
For vascular images, the network performs two segmentation predictions. The primary segmentation network, comprising the macro vessel extraction module, micro vessel enhancement module, prompt encoder module, and mask prediction module, generates the initial vessel segmentation mask. The post-processing network then generates a secondary prediction to repair disconnected vessels while preserving the intended separations. A detailed description of the implementation of each module is provided in the following sections.

\subsection{Macro Vessel Extraction Module}

The original SAM architecture consists of three main components: the image encoder, the prompt encoder, and the mask decoder. The image encoder is based on the ViT network, pre-trained with Masked Autoencoders (MAE)\cite{MAE}, and is responsible for encoding the input image. The encoded image is then combined with the prompt encoding from the prompt encoder before being passed to the mask decoder. Due to the superior receptive field of the ViT encoder, we retain the original SAM ViT encoder architecture, using it as the backbone for the macro vessel extraction module.

To enable more efficient fine-tuning, we introduce an Adapter into each ViT block while freezing the parameters of the original ViT encoder during training. The Adapter consists of two components: a feature adapter and a spatial adapter, with the latter augmented by residual connections to enhance performance. Specifically, the spatial adapter is applied with skip connections after the Multi-Head Self-Attention (MHSA) layer, followed by the addition of the feature adapter after LayerNorm. This structure allows for effective feature extraction while preserving spatial information, enabling the model to adapt to the vessel segmentation task and ensuring successful macro-vessel extraction, all while keeping the ViT encoder parameters frozen. The procedures for the spatial adapter and feature adapter are outlined as follows:

\begin{equation} S_A(X) = X + W_2 \cdot G(W_1 \cdot X) \label{eq
} \end{equation}

\begin{equation} F_A(X) = W_2 \cdot G(W_1 \cdot X) \label{eq
} \end{equation}

where \( W_1 \in \mathbb{R}^{D \times \frac{D}{4}} \), \( W_2 \in \mathbb{R}^{\frac{D}{4} \times D} \), and \( G \) denotes the GELU function.

With this design, we can fine-tune the model more efficiently while keeping the ViT encoder parameters frozen. Additionally, to improve the model's ability to extracting micro-vessel features, we incorporated a convolution-based micro vessel enhancement module.

\subsection{Micro Vessel Enhancement Module}

The micro vessel enhancement module is built on ConvNext-tiny\cite{convnext} and consists of four ConvNext stages. To enhance its ability to capture micro-vessel features, we integrate Convolutional Block Attention Module (CBAM) blocks into each ConvNext stage. This integration enriches feature representation while suppressing background noise and interference from surrounding tissues, especially in areas requiring focused attention. The four ConvNext stages contain 3, 3, 9, and 3 ConvNext blocks, respectively, resulting in feature dimensions of 96, 192, 384, and 768.

To further refine the extracted features, we introduce a feature pyramid after the ConvNext stages. This addition enhances the multi-scale representation of the convolutional features, improving their compatibility with the global features from the macro vessel extraction module. We use cross-attention to fuse the global and micro-vessel features. The micro vessel enhancement module adapts its attention based on the macro vessels, allowing it to focus more effectively on important features. Likewise, the global features extracted by the macro vessel extraction module are integrated with micro-vessel features to capture finer vascular details. This synergy enables the model to more accurately extract information from vessels of varying sizes.

\begin{figure*}[h]
\centering
\includegraphics[width=\textwidth]{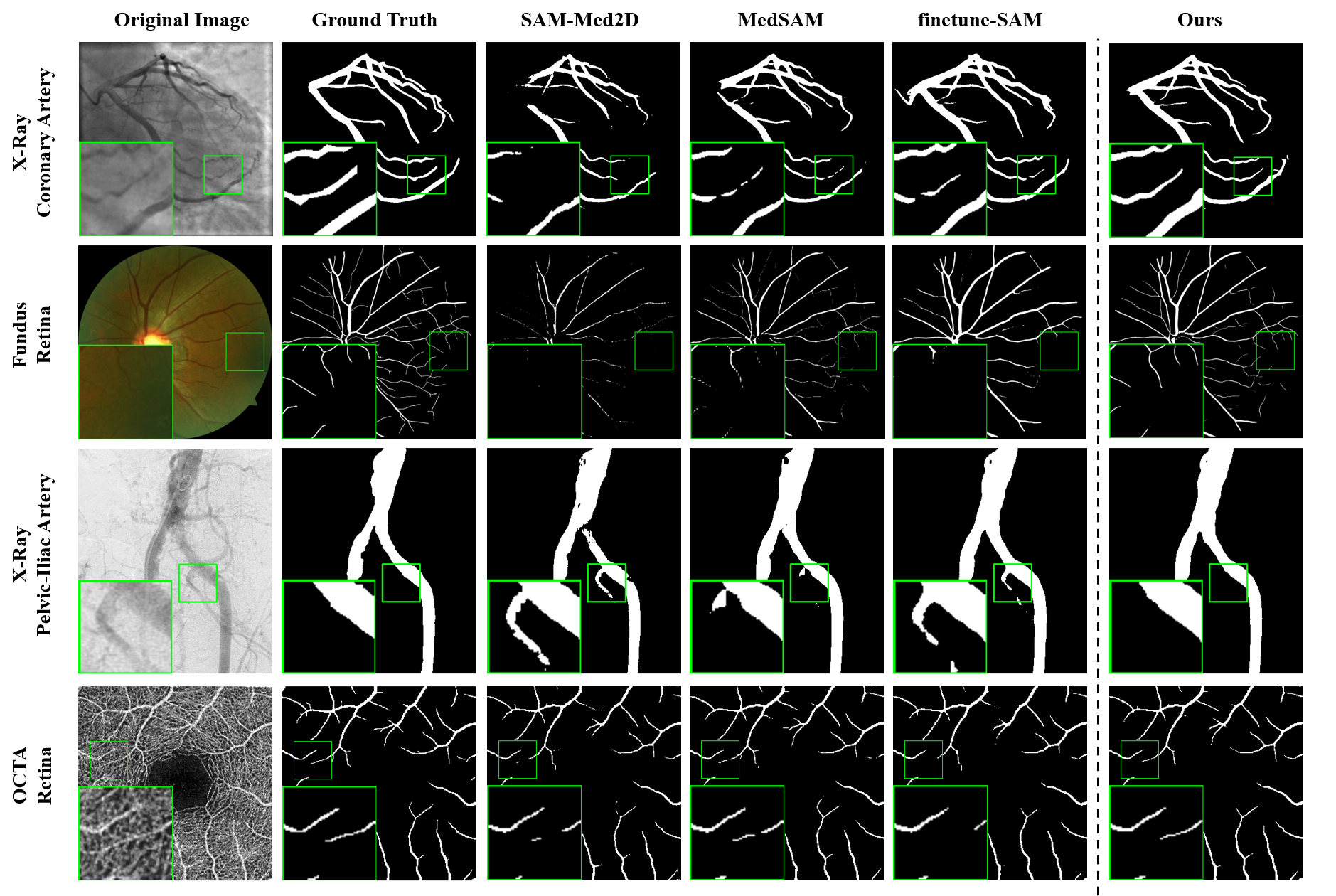}
\caption{Comparison with other SAM-based methods. From top to bottom, the images are from XCAD, ORVS, DrSAM and OCTA500-3M datasets. Due to the micro vessel enhancement module and the post-processing network, our method is better able to handle detailed information, resulting in superior performance in processing micro-vessels and maintaining vessel connectivity compared to other methods.}
\label{fig3}
\end{figure*}

\subsection{Mask Prediction Module}

The mask prediction module is based on the original mask decoder from SAM, which consists of a lightweight transformer layer and a segmentation head for mask prediction. Although this step is fundamental to segmentation, we aim to improve accuracy by incorporating ideas from prior work\cite{HQ-SAM}, which emphasize the need for both rich global semantic context and fine local boundary details.
To better integrate global and local features, we introduce a fusion of high-level and low-level features within the mask decoder. Specifically, we combine the shallow features from the first ConvNext stage with the mask features produced by the two-way transformer in the mask decoder. The shallow features capture more local information and edge details, which are crucial for accurately delineating vessel boundaries. By fusing these shallow features with the global mask features, we inject local details into the segmentation process, thereby improving the overall accuracy of the mask prediction.
This fusion approach allows the model to better preserve fine vascular details, ensuring that small vessels and complex structures are accurately segmented. The combination of high-level and low-level features helps the model achieve a more comprehensive understanding of the vascular structures, improving its segmentation performance across varying vessel sizes and shapes.

\subsection{Morphological Correction based Post-Process Network}

Unlike typical segmentation tasks, vessel segmentation imposes stringent requirements on the connectivity of the segmentation results. This connectivity is essential for accurate hemodynamic analysis, which aids in assessing patient health and prognosis, ultimately improving healthcare outcomes. To meet these requirements, we introduce a post-processing network based on morphological correction. The principle is inspired by Masked Image Modeling (MIM), but instead of masking random regions, we mask the fragmented areas of the vascular masks.

Given a network with both fragmented predicted masks and ground truth masks, the network is trained to predict the missing (i.e., fragmented) vascular structures. This allows the model to learn the correct connections at points of disconnection and repair the broken vessels. Simultaneously, for vessels that are meant to remain disconnected, the network ensures that these discontinuities are preserved. The post-processing is performed using a simple U-Net architecture, with MSE loss and ClDice loss\cite{cldice} as the loss functions. By training on a dataset that includes fragmented examples, even a basic U-Net model can learn to effectively repair disconnected vessels.

\section{Experiment}

\subsubsection{Dataset}

To train a generalizable vessel segmentation model and assess its generalization capability, we constructed the largest vessel segmentation dataset, comprising 17 publicly available or request-access datasets. These datasets cover four types of images: X-ray Coronary Artery, Fundus Retina, OCTA Retina, and X-ray Pelvic Iliac Artery. Specifically, the coronary artery dataset includes 134XCA \cite{134XCA} (also known as DCA1) and XCAD\cite{xcad}; the fundus dataset includes ARIA\cite{aria}, CHASE\_DB1\cite{chase}, DR-HAGIS\cite{drhagis}, DRIVE\cite{DRIVE}, FIVES\cite{fives}, HRF\cite{HRF}, IOSTAR\cite{iostar}, Les-AV\cite{les-av}, ORVS\cite{orvs}, and STARE\cite{stare}; the OCTA dataset includes ROSE-1\cite{OCTA-Net}, ROSSA\cite{rossa}, and OCTA500\cite{octa500}, with 39 pixel-level annotated images selected from ROSE-1; and the Pelvic-Iliac artery dataset is DrSAM\cite{drsam}.

We used 15 of these datasets for training and internal validation, each containing separate training, validation, and test sets. In addition, we evaluated the out-of-domain generalization of our model using the test sets from two additional datasets as external validation. Detailed information on the datasets is provided in Tables \ref{table1}. During training, all input images were resized to 1024×1024 and intensity normalization was applied.

\begin{table}[]
\caption{Details of datasets used in our experiments}
\begin{tabular}{cccccc}
\Xhline{1pt}
Dataset      & Modality & Resolution & Train & Val & Test \\ \hline
ARIA         & Fundus   & 768×576    & 108   & 12  & 43   \\
{CHASE\_DB1}    & Fundus   & 960×999    & 18    & 2   & 8    \\
DCA1(134XCA) & X-Ray    & 300×300    & 90    & 10  & 34   \\
DR-HAGIS     & Fundus   & 4752×3168  & 18    & 2   & 20   \\
DRIVE        & Fundus   & 584×565    & 18    & 2   & 20   \\
DrSAM        & X-Ray    & 386×448    & 360   & 40  & 100  \\
FIVE         & Fundus   & 2048×2048  & 540   & 60  & 200  \\
Les-AV       & Fundus   & 1620×1444  & 18    & 2   & 2    \\
OCTA500-3M   & OCTA     & 304×304    & 135   & 15  & 50   \\
OCTA500-6M   & OCTA     & 400×400    & 180   & 20  & 100  \\
ORVS         & Fundus   & 1444×1444  & 38    & 4   & 7    \\
ROSE-1       & OCTA     & 304×304    & 27    & 3   & 9    \\
ROSSA        & OCTA     & 320×320    & 736   &  82  & 100  \\
STARE        & Fundus   & 605×700    & 14    & 2   & 4    \\
XCAD         & X-Ray    & 512×512    & 75    & 9   & 42   \\ \hline
HRF          & Fundus   & 3504×2336  & \textbackslash     & \textbackslash   & 15   \\
IOSTAR       & Fundus   & 1024×1024  & \textbackslash     & \textbackslash   & 10   \\ \Xhline{1pt}
\label{table1}
\end{tabular}
\end{table}

\subsection{Experimental Setup}

We used the SAM model with a ViT-B architecture as the backbone, loading the corresponding pre-trained weights. In the macro vessel extraction module, we froze the ViT parameters during training, allowing only the adapters to be trained. The network and training were implemented in PyTorch on a Tesla V100-PCIE-32GB GPU.
The principal segmentation network, as the initial module in our framework, was trained for 20 epochs with a batch size of 2, a learning rate of 0.001, and the AdamW optimizer. We used a combination of binary cross-entropy and Dice loss to improve vessel detection accuracy.
The post-processing network, following the primary segmentation, was trained for 5 epochs with the same batch size, learning rate, and optimizer. Here, we applied Mean Squared Error (MSE) and ClDice loss to refine the segmentation results, ensuring more precise vessel delineation.

For evaluation metrics, we selected Dice and Intersection over Union (IoU) to assess the overall segmentation performance. Additionally, recognizing that improvements in connectivity may not significantly affect pixel-based metrics such as Dice, we introduced additional metrics to evaluate the connectivity of the segmentation results: ClDice and the normalized Betti number \( \beta_0 \). The normalized Betti number \( \beta_0 \) is calculated as follows:

\begin{equation*} \beta_0=\frac{1}{N}\displaystyle\sum_{\mathcal{i}=1}^{N} \left | CC(X_i)-CC(Y_i)\right | \label{eq} \end{equation*}

Which, referring to the methodology above, we split a mask into multiple 64×64 patches, where \(N\) represents the number of patches and \(CC\) denotes the number of disconnected components within a patch.

\subsection{Comparing with SAM Based Method}

\begin{figure}[h]
\centering
\includegraphics[width=0.5\textwidth]{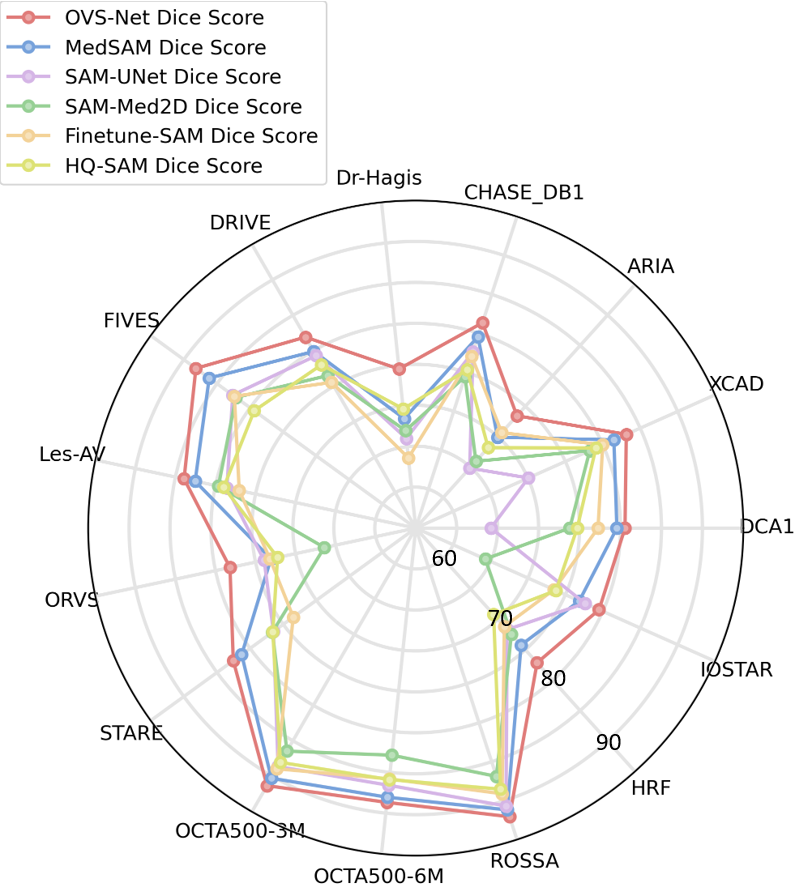}
\caption{Radar chart of scores on complex vascular datasets compared with the suboptimal SAM model.}
\label{fig4}
\end{figure}

In this section, we validate the performance of our proposed method by comparing OVS-Net with several state-of-the-art SAM-based segmentation networks, including MedSAM\cite{MedSAM}, MSA\cite{MSA}, SAM-Med2D\cite{sm2d}, Finetune-SAM\cite{ftsam}, SAM-UNet\cite{sam-unet}, and HQ-SAM\cite{HQ-SAM}, on both internal and external validation datasets. To ensure a fair comparison, we trained these SAM methods on the constructed vessel dataset, adhering to their respective default training protocols. The performance scores for different anatomical locations and modalities on both the internal and external validation datasets are presented in Table \ref{table2}.
Thanks to its multi-scale vessel extraction capability and morphological correction-based post-processing network, OVS-Net outperformed the other methods, achieving the highest average Dice, IoU, and ClDice scores, along with the lowest \( \beta_0 \) score. Compared to the second-best SAM method, OVS-Net improved the average Dice score by 1.46 and improved connectivity by 34.6\%. 
Figure \ref{fig3} visualizes the segmentation results across various datasets, showing that our method effectively captures small vessels and addresses vessel discontinuities. Figure \ref{fig4} further visualizes the radar chart comparing OVS-Net and the suboptimal SAM method on datasets with complex vascular structures, highlighting our method’s superior performance on high-resolution datasets with intricate and fine vessels (e.g., Dr-Hagis, ORVS, HRF). This demonstrates the effectiveness of OVS-Net in handling small vessel segmentation, which is crucial for clinical applications where precise vessel delineation impacts diagnostic and therapeutic decisions.
Furthermore, OVS-Net maintained its competitive edge on the external dataset, achieving the best performance with a 1.79 improvement in average Dice scores and a 30.7\% improvement in connectivity. These results not only demonstrate the effectiveness of OVS-Net but also highlight its strong generalization ability across diverse datasets and imaging modalities, underscoring its potential for broader clinical applications.


\begin{table*}[]
\centering
\caption{Compared with SAM-based methods, we report the individual scores for coronary artery, fundus, OCTA, and pelvic-iliac artery in the internal validation dataset, as well as the scores in the external validation dataset. The best scores are in bold, and the second-best methods are underlined; the same applies below.}
\begin{tabular}{ccccccccc}
\Xhline{1pt}
\multirow{2}{*}{Method}                                                      & \multicolumn{4}{c}{Xray-Coronary\_Artery}                               & \multicolumn{4}{c}{Fundus-Retina}                                       \\ \cline{2-9}
                                                                             & Dice↑          & IoU↑           & clDice↑        & \(\beta_0\)↓  & Dice↑          & IoU↑           & clDice↑        & \(\beta_0\)↓  \\ \hline
MedSAM\cite{MedSAM}                                                                       & 80.40\scalebox{0.8}{±3.78}     & 67.38\scalebox{0.8}{±5.28}     & 86.59\scalebox{0.8}{±4.04}     & 0.26\scalebox{0.8}{±0.12}     & 82.57\scalebox{0.8}{±9.55}     & 71.30\scalebox{0.8}{±12.24}    & 82.97\scalebox{0.8}{±10.52}    & 0.33\scalebox{0.8}{±0.39}     \\
MSA\cite{MSA}                                                                          & 74.60\scalebox{0.8}{±3.24}          & 59.60\scalebox{0.8}{±4.09}          & 81.44\scalebox{0.8}{±4.80}          & 0.52\scalebox{0.8}{±0.25}          & 71.72\scalebox{0.8}{±8.58}          & 56.49\scalebox{0.8}{±8.92}          & 71.24\scalebox{0.8}{±9.44}          & 0.44\scalebox{0.8}{±0.49}          \\
SAM-Med2D\cite{sm2d}                                                                    & 75.74\scalebox{0.8}{±6.02}          & 61.32\scalebox{0.8}{±7.72}          & 80.41\scalebox{0.8}{±6.94}          & 0.50\scalebox{0.8}{±0.27}          & 78.43\scalebox{0.8}{±12.66}          & 66.02\scalebox{0.8}{±14.67}          & 78.98\scalebox{0.8}{±13.36}          & 0.35\scalebox{0.8}{±0.38}          \\
Finetune-SAM\cite{ftsam}                                                                 & 78.44\scalebox{0.8}{±4.48}          & 64.75\scalebox{0.8}{±6.07}          & 84.22\scalebox{0.8}{±5.21}          & 0.31\scalebox{0.8}{±0.13}          & 79.13\scalebox{0.8}{±9.78}          & 66.37\scalebox{0.8}{±11.38}          & 77.61\scalebox{0.8}{±11.04}          & 0.37\scalebox{0.8}{±0.43}          \\
SAM-UNet\cite{sam-unet}                                                                     & 67.09\scalebox{0.8}{±22.82}          & 43.11\scalebox{0.8}{±19.99}          & 61.61\scalebox{0.8}{±23.63}          & 0.55\scalebox{0.8}{±0.36}          & 79.33\scalebox{0.8}{±13.39}          & 67.36\scalebox{0.8}{±14.49}          & 79.61\scalebox{0.8}{±14.55}          & 0.28\scalebox{0.8}{±0.29}          \\
HQ-SAM\cite{HQ-SAM}                                                                       & 76.61\scalebox{0.8}{±4.30}          & 62.28\scalebox{0.8}{±5.63}          & 80.42\scalebox{0.8}{±5.45}          & 0.77\scalebox{0.8}{±0.45}          & 77.37\scalebox{0.8}{±10.70}          & 64.15\scalebox{0.8}{±12.40}          & 77.40\scalebox{0.8}{±11.08}          & 0.39\scalebox{0.8}{±0.31}          \\ \hline
OVS-Net & \textbf{81.67\scalebox{0.8}{±3.50}}    & \textbf{ 69.17\scalebox{0.8}{±5.03}}    & \textbf{87.37\scalebox{0.8}{±4.12}} & \textbf{0.20\scalebox{0.8}{±0.09}} & \textbf{85.06\scalebox{0.8}{±8.51}}    & \textbf{74.83\scalebox{0.8}{±11.35}}    & \textbf{87.07\scalebox{0.8}{±8.85}} & \textbf{0.18\scalebox{0.8}{±0.20}} \\ \Xhline{1pt}
\multirow{2}{*}{Method}                                                      & \multicolumn{4}{c}{OCTA-Retina}                                         & \multicolumn{4}{c}{Xray-Pelvic\_Iliac\_Artery}                          \\ \cline{2-9}
                                                                             & Dice↑          & IoU↑           & clDice↑        & \(\beta_0\)↓  & Dice↑          & IoU↑           & clDice↑        & \(\beta_0\)↓  \\ \hline
MedSAM\cite{MedSAM}                                                                       & 89.58\scalebox{0.8}{±3.19}    & 81.28\scalebox{0.8}{±5.07}    & 91.45\scalebox{0.8}{±3.27} & 0.98\scalebox{0.8}{±0.38}    &  96.62\scalebox{0.8}{±1.64}    & 93.51\scalebox{0.8}{±2.97}    & 97.56\scalebox{0.8}{±2.59} & 0.05\scalebox{0.8}{±0.05}    \\
MSA\cite{MSA}                                                                          & 80.35\scalebox{0.8}{±6.32}          & 67.59\scalebox{0.8}{±8.41}          & 83.14\scalebox{0.8}{±6.32}          & 1.68\scalebox{0.8}{±0.57}          & 94.88\scalebox{0.8}{±2.33}          & 90.35\scalebox{0.8}{±4.09}          & 96.14\scalebox{0.8}{±3.96}          & 0.10\scalebox{0.8}{±0.08}          \\
SAM-Med2D\cite{sm2d}                                                                    & 84.99\scalebox{0.8}{±3.73}          & 74.08\scalebox{0.8}{±5.49}          & 87.50\scalebox{0.8}{±4.16}          & 1.70\scalebox{0.8}{±0.54}          & 95.14\scalebox{0.8}{±2.23}          & 90.82\scalebox{0.8}{±3.92}          & 96.40\scalebox{0.8}{±3.30}          & 0.11\scalebox{0.8}{±0.08}          \\
Finetune-SAM\cite{ftsam}                                                                 & 87.41\scalebox{0.8}{±3.72}          & 77.81\scalebox{0.8}{±5.61}          & 89.39\scalebox{0.8}{±4.00}          & 1.20\scalebox{0.8}{±0.42}          & 96.31\scalebox{0.8}{±1.85}          & 92.94\scalebox{0.8}{±3.33}          & 97.31\scalebox{0.8}{±2.82}          & 0.05\scalebox{0.8}{±0.42}          \\
SAM-UNet\cite{sam-unet}                                                                     & 88.25\scalebox{0.8}{±3.97}          & 79.19\scalebox{0.8}{±6.12}          & 90.49\scalebox{0.8}{±3.98}          & 1.05\scalebox{0.8}{±0.37}          & 80.95\scalebox{0.8}{±19.25}          & 71.66\scalebox{0.8}{±22.97}          & 82.22\scalebox{0.8}{±18.71}          & 0.21\scalebox{0.8}{±0.17}          \\
HQ-SAM\cite{HQ-SAM}                                                                       & 86.76\scalebox{0.8}{±4.53}          & 76.82\scalebox{0.8}{±6.47}          & 88.72\scalebox{0.8}{±4.76}          & 1.93\scalebox{0.8}{±0.57}          & 95.10\scalebox{0.8}{±2.26}          & 90.73\scalebox{0.8}{±3.98}          & 93.84\scalebox{0.8}{±4.89}          & 0.2861\scalebox{0.8}{±0.24}        \\ \hline
OVS-Net & \textbf{90.36\scalebox{0.8}{±3.21}} & \textbf{82.57\scalebox{0.8}{±5.18}} & \textbf{92.20\scalebox{0.8}{±3.08}} & \textbf{0.67\scalebox{0.8}{±0.30}} & \textbf{96.64\scalebox{0.8}{±1.90}} & \textbf{93.56\scalebox{0.8}{±3.43}} & \textbf{97.57\scalebox{0.8}{±2.94}} & \textbf{0.05\scalebox{0.8}{±0.61}} \\ \Xhline{1pt}
\multirow{2}{*}{Method}                                                      & \multicolumn{4}{c}{External Validation Dataset}                             & \multicolumn{4}{c}{Average}                                      \\ \cline{2-9}
                                                                             & Dice↑          & IoU↑           & clDice↑        & \(\beta_0\)↓  & Dice↑          & IoU↑           & clDice↑        & \(\beta_0\)↓  \\ \hline
MedSAM\cite{MedSAM}                                                                       & 74.83\scalebox{0.8}{±3.29}    &  59.89\scalebox{0.8}{±4.24}    & 73.16\scalebox{0.8}{±6.25} & 0.40\scalebox{0.8}{±0.14}    & 86.61\scalebox{0.8}{±8.47}         & 77.28\scalebox{0.8}{±12.03}          & 88.06\scalebox{0.8}{±9.18}          & 0.52\scalebox{0.8}{±0.49}          \\
MSA\cite{MSA}                                                                          & 71.07\scalebox{0.8}{±3.51}          & 55.24\scalebox{0.8}{±4.09}          & 63.85\scalebox{0.8}{±4.96}          & 0.47\scalebox{0.8}{±0.15}          & 78.22\scalebox{0.8}{±10.20}          & 65.35\scalebox{0.8}{±13.59}          & 79.57\scalebox{0.8}{±11.44}          & 0.85\scalebox{0.8}{±0.79}          \\
SAM-Med2D\cite{sm2d}                                                                    & 69.50\scalebox{0.8}{±7.59}          & 53.73\scalebox{0.8}{±8.28}          & 68.54\scalebox{0.8}{±6.87}          & 0.55\scalebox{0.8}{±0.38}          & 82.60\scalebox{0.8}{±10.61}          & 71.57\scalebox{0.8}{±13.69}          & 84.25\scalebox{0.8}{±11.24}          & 0.82\scalebox{0.8}{±0.79}          \\
Finetune-SAM\cite{ftsam}                                                                 & 71.76\scalebox{0.8}{±3.34}          & 56.06\scalebox{0.8}{±4.14}          & 68.69\scalebox{0.8}{±5.57}          & 0.43\scalebox{0.8}{±0.14}          & 84.19\scalebox{0.8}{±9.27}          & 73.69\scalebox{0.8}{±12.68}          & 84.84\scalebox{0.8}{±10.70}          & 0.63\scalebox{0.8}{±0.58}          \\
SAM-UNet\cite{sam-unet}                                                                     & 73.40\scalebox{0.8}{±4.34}          & 58.17\scalebox{0.8}{±5.50}          & 72.92\scalebox{0.8}{±7.82}          & 0.39\scalebox{0.8}{±0.10}          & 80.73\scalebox{0.8}{±15.48}          & 69.91\scalebox{0.8}{±17.45}          & 82.18\scalebox{0.8}{±15.81}          & 0.57\scalebox{0.8}{±0.48}          \\
HQ-SAM\cite{HQ-SAM}                                                                       & 70.47\scalebox{0.8}{±4.86}          & 54.63\scalebox{0.8}{±5.87}          & 73.35\scalebox{0.8}{±4.50}          & 0.57\scalebox{0.8}{±0.43}          & 82.91\scalebox{0.8}{±10.03}          & 71.94\scalebox{0.8}{±13.36}          & 83.86\scalebox{0.8}{±10.26}          & 0.97\scalebox{0.8}{±0.85}          \\ \hline
OVS-Net & \textbf{77.62\scalebox{0.8}{±3.95}} & \textbf{63.60\scalebox{0.8}{±5.32}} & \textbf{78.93\scalebox{0.8}{±5.80}} & \textbf{0.27\scalebox{0.8}{±0.05}} & \textbf{88.05\scalebox{0.8}{±7.49}} & \textbf{79.38\scalebox{0.8}{±10.95}} & \textbf{90.14\scalebox{0.8}{±7.40}} & \textbf{0.34\scalebox{0.8}{±0.33}} \\ \Xhline{1pt}
\end{tabular}
\label{table2}
\end{table*}

\subsection{Comparing with Task-Specific Expert Methods}

\begin{figure*}[!t]
\centering
\includegraphics[width=\textwidth]{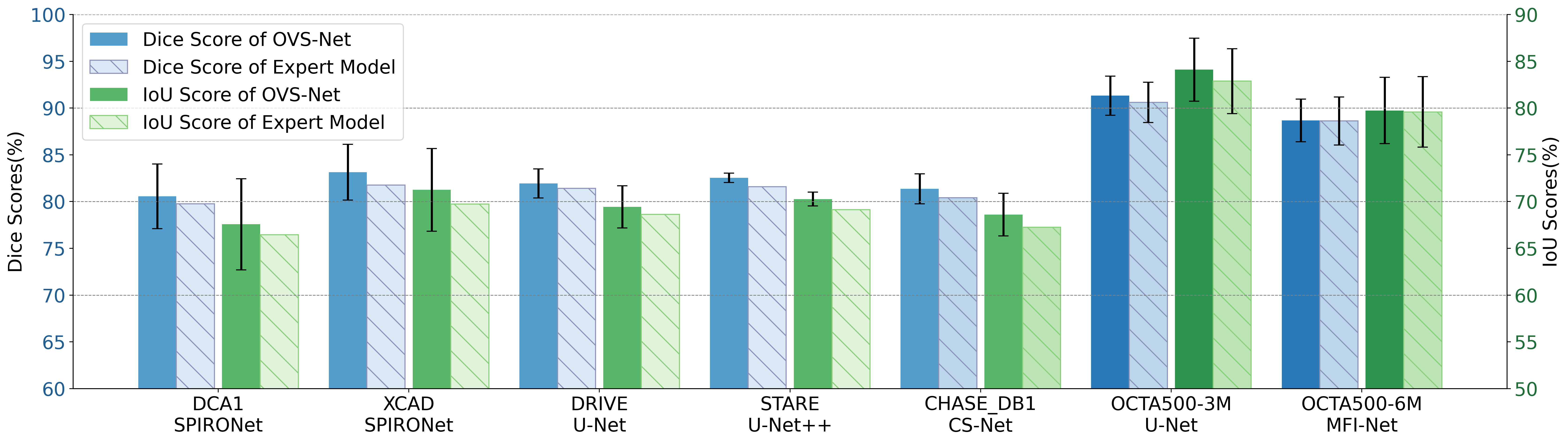}
\caption{Bar chart comparing the scores of OVS-Net and the expert model. Here we use the standard deviation as the error bar standard. 
For expert model selection, we select the expert model with the highest Dice score to plot the bar chart.}
\label{fig5}
\end{figure*}

In this section, we present a comprehensive quantitative comparison of our method against various expert models on both internal and external validation datasets, providing a clear perspective on its performance relative to established benchmarks. The score comparisons for representative datasets are shown in Tables \ref{table3}, \ref{table4}, and \ref{table5}, while external validation dataset scores are provided in Table \ref{table6}. The scores of expert models are taken from \cite{spironet}, \cite{fundusscore}, \cite{octscore}, \cite{hrfscore1}, \cite{bionet}, and \cite{iostar}, with \cite{octscore} being the only source reporting standard deviation. We also plotted a bar chart in Figure \ref{fig5} to compare the scores of OVS-Net with the expert models. For the expert model selection, we chose the model with the highest Dice score for comparison.

In Table \ref{table3}, we compare the segmentation scores of our method with 11 prominent expert models, including U-Net\cite{U-Net}, U-Net++\cite{unet++}, Attention-Unet\cite{attention-unet}, CE-Net\cite{cenet}, CAU-Net\cite{caunet}, \(\text{CS}^2\)-Net\cite{cs2net}, FR-Net\cite{FR-Net}, DE-DCGCN-EE\cite{DE-DCGCN-EE}, GT-DLA-dsHFF\cite{GT-DLA-dsHFF}, and SPIRONet\cite{spironet}, across two coronary artery datasets. Our method outperforms all others in coronary artery segmentation, achieving average Dice scores of 80.55 and 83.14, surpassing the latest state-of-the-art methods. This notable improvement underscores the effectiveness of our model in accurately delineating vascular structures, which is crucial for clinical decision-making in cardiovascular health.

\begin{table}[]
\caption{Quantitative comparison of our OVS-Net and SOTA task-specific expert methods on coronary artery segmentation task}
\setlength{\tabcolsep}{5pt}
\centering
\begin{tabular}{ccccc}
\Xhline{1pt}
\multirow{2}{*}{Method} & \multicolumn{2}{c}{DCA1(134XCA)}                                                & \multicolumn{2}{c}{XCAD}                                                \\
   \cline{2-5}                     & Dice                                 & IoU                                & Dice                                 & IoU                                \\ \hline
U-Net\cite{U-Net}                   & 78.90                              & 65.31                              & 80.74                              & 67.98                              \\
U-Net++\cite{unet++}                 & 78.45                              & 64.69                              & 80.43                              & 67.56                              \\
Attention-Unet\cite{attention-unet}          & 78.04                              & 64.18                              & 79.98                              & 66.94                              \\
CE-Net\cite{cenet}                  & 77.84                              & 63.87                              & 79.95                              & 66.8                               \\
CAU-Net\cite{caunet}                 & 77.82                              & 63.88                              & 79.31                              & 66.06                              \\
TransUNet\cite{transunet}               & 78.82                              & 65.19                              & 80.24                              & 67.28                              \\
\text{CS}$^2$\text{-Net}\cite{cs2net}                 & 77.87                              & 63.94                              & 79.23                              & 65.98                              \\
FR-UNet\cite{FR-UNet}                 & 79.59                              & 66.22                              & 79.79                              & 66.66                              \\
DE-DCGCN-EE\cite{DE-DCGCN-EE}             & 77.82                              & 63.87                              & 79.12                              & 65.76                              \\
GT-DLA-dsHFF\cite{GT-DLA-dsHFF}            & 77.17                              & 62.97                              & 80.44                              & 67.53                              \\
SPIRONet\cite{spironet}                & 79.75                              & 66.45                              & 81.76                              & 69.73                              \\ \hline
OVS-Net  & \multicolumn{1}{c}{\textbf{\makecell{80.55\\\hspace*{\fill}\scriptsize{   ±3.46}}}} & \multicolumn{1}{c}{\textbf{\makecell{67.57\\\hspace*{\fill}\scriptsize{   ±4.88}}}} & \multicolumn{1}{c}{\textbf{\makecell{83.14\\\hspace*{\fill}\scriptsize{   ±2.98}}}} & \multicolumn{1}{c}{\textbf{\makecell{71.25\\\hspace*{\fill}\scriptsize{   ±4.43}}}} \\ \Xhline{1pt}
\label{table3}
\end{tabular}
\end{table}

For retinal fundus datasets, Table \ref{table4} compares our approach with five expert models: U-Net\cite{U-Net}, U-Net++\cite{unet++}, Attention-UNet\cite{attention-unet}, CS-Net\cite{cs-net}, and H2Former\cite{h2former} across three datasets. On all three fundus datasets, our method consistently achieves the highest average Dice and IoU scores, demonstrating its robustness in handling variations in retinal images. While OVS-Net achieves the highest Dice score on the DRIVE dataset, it records a slightly lower IoU compared to CS-Net.

\begin{table}[]
\caption{Quantitative comparison of our OVS-Net and SOTA task-specific expert methods on Fundus segmentation task}
\resizebox{0.5\textwidth}{!}{
\begin{tabular}{ccccccc}
\Xhline{1pt}
\multirow{2}{*}{Method}                                                      & \multicolumn{2}{c}{DRIVE}       & \multicolumn{2}{c}{STARE}       & \multicolumn{2}{c}{CHASE\_DB1}  \\ \cline{2-7}
                                                                             & Dice           & IoU            & Dice           & IoU            & Dice           & IoU            \\ \hline
U-Net\cite{U-Net}                                                                        & 81.41          & 68.64          & 81.18          & 68.56          & 78.98          & 65.26          \\
U-Net++\cite{unet++}                                                                      & 81.14          & 68.27          & 81.50          & 69.02          & 80.15          & 66.88          \\
Attention-Unet\cite{attention-unet}                                                               & 80.39          & 67.21          & 81.06          & 68.39          & 79.64          & 66.17          \\
CS-Net\cite{cs-net}                                                                       & 80.39          & \textbf{70.17} & 81.59          & 69.12          & 80.42          & 67.25          \\
H2Former\cite{h2former}                                                                     & 74.54          & 59.71          & 77.56          & 63.42          & 74.92          & 59.98          \\ \hline
OVS-Net & {\textbf{\makecell{81.93\\\hspace*{\fill}\scriptsize{   ±1.55}}}} & {\makecell{69.43\\\hspace*{\fill}\scriptsize{   ±2.25}}}    & {\textbf{\makecell{82.54\\\hspace*{\fill}\scriptsize{   ±0.50}}}} & {\textbf{\makecell{70.27\\\hspace*{\fill}\scriptsize{   ±0.73}}}} & {\textbf{\makecell{81.36\\\hspace*{\fill}\scriptsize{   ±1.61}}}} & {\textbf{\makecell{68.61\\\hspace*{\fill}\scriptsize{   ±2.28}}}} \\ \Xhline{1pt}
\label{table4}
\end{tabular}}
\end{table}

\begin{table}[h]
\centering
\caption{Quantitative comparison of our OVS-Net and SOTA task-specific expert methods on OCTA segmentation task}
\setlength{\tabcolsep}{5pt}
\begin{tabular}{ccccccc}
\Xhline{1pt}
\multirow{2}{*}{Method}                                                      & \multicolumn{2}{c}{OCTA500-3M}  & \multicolumn{2}{c}{OCTA500-6M}  \\ \cline{2-5}
                                                                             & Dice           & IoU            & Dice           & IoU            \\ \hline
U-Net\cite{U-Net}                                                                        & 90.60\scalebox{0.8}{±2.16}          & 82.88\scalebox{0.8}{±3.47}          & 88.28\scalebox{0.8}{±2.59}          & 79.11\scalebox{0.8}{±3.97}          \\
MFI-Net\cite{mfi}                                                                      & 90.15\scalebox{0.8}{±1.59}          & 82.15\scalebox{0.8}{±2.57}          & 88.62\scalebox{0.8}{±2.57}          & 79.59\scalebox{0.8}{±3.77}          \\
Swin-Unet\cite{swin-unet}                                                                    & 84.42\scalebox{0.8}{±2.87}          & 73.14\scalebox{0.8}{±4.19}          & 82.40\scalebox{0.8}{±2.68}          & 70.16\scalebox{0.8}{±3.78}          \\
H2Former\cite{h2former}                                                                     & 83.82\scalebox{0.8}{±1.86}          & 72.24\scalebox{0.8}{±3.22}          & 82.45\scalebox{0.8}{±2.57}          & 70.25\scalebox{0.8}{±3.74}          \\ \hline
OVS-Net& \textbf{91.32\scalebox{0.8}{±2.10}} & \textbf{84.10\scalebox{0.8}{±3.38}} & {\textbf{88.68\scalebox{0.8}{±2.28}}}    & {\textbf{79.74\scalebox{0.8}{±3.55}}}    \\ \Xhline{1pt}
\label{table5}
\end{tabular}
\end{table}

\begin{table}[]
\centering
\caption{Quantitative score comparison with expert models on two external validation datasets}
\begin{tabular}{c@{\hspace{5pt}}cc@{\hspace{5pt}}c}
\Xhline{1pt}
Method                                                                       & HRF                                & Method                                                                       & IOSTAR                             \\ \hline
U-Net\cite{U-Net}                                                                        & 72.39                              & U-Net\cite{U-Net}                                                                        & 76.96                              \\
CS-Net\cite{cs-net}                                                                       & 71.04                              & DNN\cite{dnn}                                                                          & 76.01                              \\
SA-UNet\cite{sa-unet}                                                                      & 71.18                              & R2UNet\cite{r2unet}                                                                       & 77.16                        \\
SCS-Net\cite{SCS-Net}                                                                      & 70.66                              & Symmetry Filter\cite{symmetry}                                                              & 74.31                              \\
CogSeg\cite{cognet}                                                                       & 74.75                              & VA-UFL\cite{va-ufl}                                                                       & 73.99                              \\
SuperVessel\cite{supervessel}                                                                  & 76.74                        & MSR-Net\cite{msrnet}                                                                      & 76.17                              \\
BioNet\cite{bionet}                                                                       & 76.26                              & SUD-GAN\cite{sud-gan}                                                                      & 76.96                              \\ \hline
OVS-Net & {\textbf{\makecell{77.14\\\hspace*{\fill}\scriptsize{   ±4.05}}}} & OVS-Net & \textbf{\makecell{78.82\\\hspace*{\fill}\scriptsize{   ±3.43}}} \\ \Xhline{1pt}
\end{tabular}
\label{table6}
\end{table}

\begin{table}[]
\caption{Comparison of connectivity-based post-processing methods. We conducted a comparison of our method with connectivity-focused post-processing approaches on the DRIVE, CHASE\_DB1, and DCA1 datasets.}
\resizebox{0.5\textwidth}{!}{
\begin{tabular}{c@{\hspace{3pt}}c@{\hspace{3pt}}c@{\hspace{6pt}}c@{\hspace{3pt}}c@{\hspace{6pt}}c@{\hspace{3pt}}c}
\Xhline{1pt}
\multirow{2}{*}{Method}                                                      & \multicolumn{2}{c}{DRIVE}     & \multicolumn{2}{c}{CHASE\_DB1} & \multicolumn{2}{c}{DCA1}   \\ \cline{2-7}
                                                                             & clDice        & \(\beta_0\)            & clDice         & \(\beta_0\)            & clDice        & \(\beta_0\)         \\ \hline
UNet\cite{U-Net}  & 81.9\scalebox{0.8}{±4.0}   & 
0.79\scalebox{0.8}{±0.16}  & 
84.1\scalebox{0.8}{±2.6}  & 
0.40\scalebox{0.8}{±0.13}  & 
84.5\scalebox{0.8}{±4.6}  & 
0.36\scalebox{0.8}{±0.17} \\
ErrorNet\cite{errornet}  & 81.6\scalebox{0.8}{±3.9}  & 
0.85\scalebox{0.8}{±0.18}  & 
83.7\scalebox{0.8}{±2.7}  & 
0.41\scalebox{0.8}{±0.10}  & 
85.5\scalebox{0.8}{±4.5}  & 
0.27\scalebox{0.8}{±0.13}  \\
TopoLoss\cite{topoloss}   & 81.9\scalebox{0.8}{±3.8}  & 
0.77\scalebox{0.8}{±0.14}  & 
84.4\scalebox{0.8}{±2.5}  & 
0.40\scalebox{0.8}{±0.06}  & 
84.9\scalebox{0.8}{±4.8}  & 
0.28\scalebox{0.8}{±0.11}  \\  
LIOT\cite{LIOT}   & 81.5\scalebox{0.8}{±4.3}          & 
0.80\scalebox{0.8}{±0.16}          & 
83.9\scalebox{0.8}{±2.7}           & 
0.41\scalebox{0.8}{±0.12}          & 
85.1\scalebox{0.8}{±4.4}          & 
0.33\scalebox{0.8}{±0.12}       \\
DconnNet\cite{DconnNet}   & 81.5\scalebox{0.8}{±4.3}          & 
0.80\scalebox{0.8}{±0.16}          & 
83.9\scalebox{0.8}{±2.7}           & 
0.41\scalebox{0.8}{±0.12}          & 
85.1\scalebox{0.8}{±4.4}          & 
0.33\scalebox{0.8}{±0.12}       \\ \hline
OVS-Net& \textbf{82.1\scalebox{0.8}{±3.90}} & \textbf{0.61\scalebox{0.8}{±0.13}}   & 
\textbf{84.4\scalebox{0.8}{±2.3}}  & 
\textbf{0.26\scalebox{0.8}{±0.03}} & 
\textbf{87.4\scalebox{0.8}{±4.4}} & 
\textbf{0.22\scalebox{0.8}{±0.11}} \\ \Xhline{1pt}
\label{table7}
\end{tabular}}
\end{table}

Table \ref{table5} presents a detailed comparison of our method on two retinal OCTA datasets against 4 expert models, including U-Net\cite{U-Net}, MFI-Net\cite{mfi}, Swin-Unet\cite{swin-unet} and H2Former\cite{h2former}. Our method achieves the highest Dice and IoU scores on the OCTA500-3M and OCTA500-6M dataset, showcasing its effectiveness in segmenting intricate vascular patterns typical in OCTA images further affirming our model’s versatility and capability to perform comparably to, or even surpass, expert models across multiple tasks.

\begin{figure*}[!t]
\centering
\includegraphics[width=\textwidth]{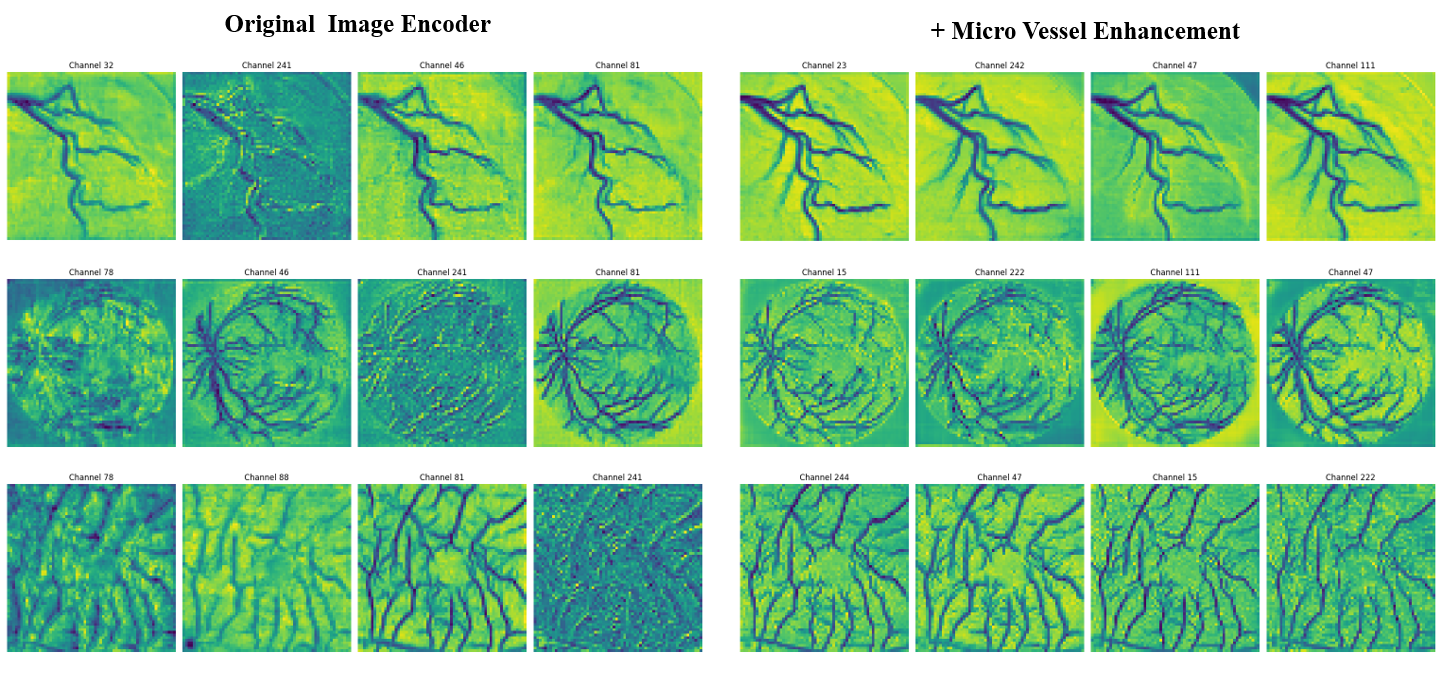}
\caption{Feature map visualization. We selected the four channels with the highest average activation values for visualization. The feature maps after incorporating the micro vessel enhancement module show notable improvement in extracting vascular network features.}
\label{fig6}
\end{figure*}

In Table \ref{table6}, we present a comparison of Dice scores of OVS-Net on two external validation datasets, HRF and IOSTAR, compared to 13 expert models, including U-Net\cite{U-Net}, CS-Net\cite{cs-net}, SAUNet\cite{sa-unet}, SCS-Net\cite{SCS-Net}, CogSeg\cite{cognet}, SuperVessel\cite{supervessel}, BioNet\cite{bionet}, DNN\cite{dnn}, R2UNet\cite{r2unet}, Symmetry Filter\cite{symmetry}, VA-UFL\cite{va-ufl}, MSR-Net\cite{msrnet}, and SUD-GAN\cite{sud-gan}. Our method achieved the highest scores on both datasets, with an average Dice score of 77.14 on the HRF dataset and 79.19 on the IOSTAR dataset. This demonstrates the potential of our approach to address the limitations of task-specific expert models, showing its ability to effectively recognize vascular structures of varying scales across different data distribution scenarios. Such a capability is particularly valuable for clinical applications, as expert models often struggle to generalize across different imaging conditions or data variations. Additionally, they may require extensive retraining on data obtained from different imaging modalities, scanners, or protocols, leading to substantial operational costs.

Additionally, we provide a comparison of connectivity metrics with expert models in Table \ref{table7}. This table reports connectivity performance on the DRIVE, CHASE, and DCA1 datasets against specific segmentation networks or loss functions: Unet\cite{U-Net}, ErrorNet\cite{errornet}, TopoLoss\cite{topoloss}, LIOT\cite{LIOT}, and DconnNet\cite{DconnNet}. The expert model scores are reported from \cite{deepclosing}. As shown in the table, our method, enhanced by a morphological correction-based post-processing network, achieves leading performance compared to specific expert models. This result is particularly notable, as it highlights our approach's ability to not only deliver high-quality segmentation but also ensure that the segmented vessels maintain structural continuity—a crucial factor in clinical applications where accurate vascular connectivity is essential for diagnosis and treatment.

\begin{table}[]
\centering
\caption{Quantitative Comparison of Ablation Study for OVS-Net}
\begin{tabular}{cccccc}
\Xhline{1pt}
Adapter & \begin{tabular}[c]{@{}c@{}}Micro Vessel\\ Enhancement\end{tabular} & \begin{tabular}[c]{@{}c@{}}Feature\\ Connection\end{tabular} & \begin{tabular}[c]{@{}c@{}}Post-\\ processing\end{tabular} & Dice           & \(\beta_0\)       \\ \hline
×       & ×                                                                  & ×                                                            & ×                                                          & 86.97          & 0.53          \\
\checkmark       & ×                                                                  & ×                                                            & ×                                                          & 87.21          & 0.53          \\
\checkmark       & \checkmark                                                                  & ×                                                            & ×                                                          & 88.24          & 0.49          \\
\checkmark       & \checkmark                                                                  & \checkmark                                                            & ×                                                          & 88.28          & 0.48          \\
\checkmark       & \checkmark                                                                  & \checkmark                                                            & \checkmark                                                          & \textbf{88.37} & \textbf{0.34} \\ \Xhline{1pt}
\end{tabular}
\label{table8}
\end{table}

\subsection{Ablation Study}

To validate the effectiveness of the proposed improvements, we conducted a series of ablation studies on the internal validation dataset. In these experiments, we incrementally added the adapter, micro vessel enhancement module, feature concatenation in the Mask Decoder, and the morphological correction-based post-processing network to the baseline model. The Dice score and Betti number \(\beta_0\)  were used as performance metrics, and the results for each configuration are reported in Table \ref{table8}. The experimental results show that each modification improves SAM's performance on vascular imaging, with the micro vessel enhancement module and post-processing network having the most significant impact on segmentation accuracy.

\begin{figure}[]
\centering
\includegraphics[width=0.5\textwidth]{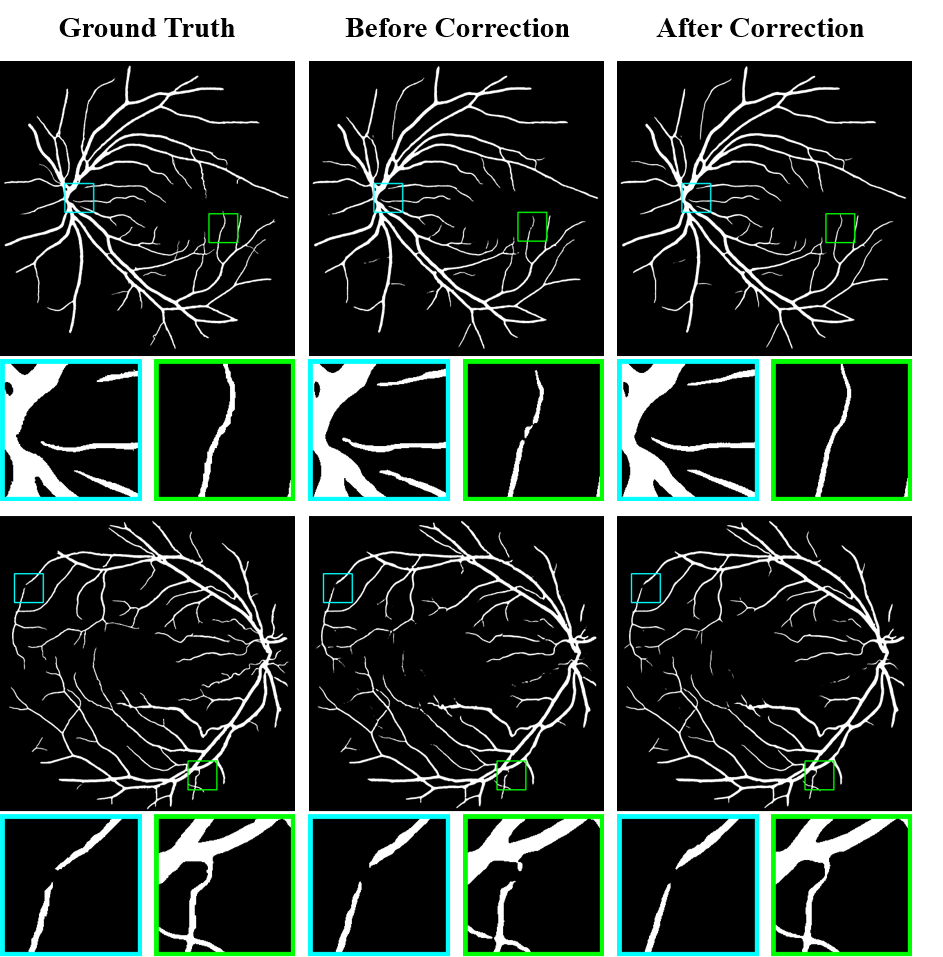}
\caption{Post-processing effect visualization. The blue boxes indicate preserved original discontinuities, while the green boxes represent vessels whose predicted discontinuities have been successfully repaired.}
\label{fig7}
\end{figure}

After integrating the micro vessel enhancement module, we observed a 1.31-point increase in the average Dice score, indicating a significant improvement in segmentation precision. To better illustrate this improvement, we visualized the feature activation maps before and after incorporating the module, as shown in Figure \ref{fig6}. In this visualization, we selected the four channels with the highest mean activation for closer examination. The feature maps show a clear enhancement in delineating vascular boundaries and a stronger focus on small vessel after the module was added. This suggests that the micro vessel enhancement module helps the network capture features more effectively across different vessel sizes, which is crucial for accurate segmentation in complex vascular structures.

Furthermore, after incorporating the post-processing network, we observed a reduction of 0.14 in the Betti number \(\beta_0\), indicating improved connectivity in the segmentation results. To visually demonstrate this, Figure \ref{fig7} compares the segmentation outcomes from the Primary Segmentation phase, before and after applying the post-processing network. In this figure, blue boxes highlight areas where the network successfully preserved inherent disconnections in the ground truth, while green boxes show where it effectively corrected segmentation breaks. Importantly, the post-processing network not only repairs these breaks but also preserves natural disconnections found in the original data, which is valuable for clinical applications as it ensures the segmented results align with the anatomical features of vascular structures.

\section{Conclusion}

In this paper, we propose OVS-Net, a novel network exploring the Segment Anything Model (SAM) for optimized vessel segmentation. OVS-Net features five key modules: a macro vessel extraction module, micro vessel enhancement module, prompt encoder, mask prediction module, and a morphological correction-based post-processing network.
The hybrid encoder architecture exploits SAM’s Vision Transformer (ViT) for macro feature extraction with a convolution-based module for enhancing fine vessel structures and edges, critical for precise segmentation. Low-level feature fusion in the mask prediction module integrates global and local features, enabling accurate segmentation across varying scales and complexities. The post-processing network further improves clinical relevance by correcting vessel disconnections while preserving natural anatomical gaps, ensuring segmentation results are both accurate and meaningful.

Comprehensive experiments across 17 datasets demonstrate that OVS-Net achieves superior segmentation accuracy, enhanced connectivity, and robust generalization across diverse imaging modalities. While this study focuses on technical evaluation, future work will aim to integrate OVS-Net into clinical workflows, conducting user studies to assess its practical impact and potential to improve diagnostic and therapeutic outcomes, ultimately benefiting patient care.

% \printbibliography
% \bibliographystyle{IEEEtran}  % 引用样式
% \bibliography{reference}      % 文献库

\bibliographystyle{IEEEtran}
% \small\bibliography{reference}

% \newpage

% If you have an EPS/PDF photo (graphics package needed), extra braces are
 

% \end{IEEEbiography}


% \begin{IEEEbiographynophoto}{Dongning Song} received the B.E. degree from the School of Automation, Central South University, in 2024, where he is currently pursuing the doctor’s degree. His research interests include medical image analysis and image segmentation

\vfill

\end{document}